\documentclass[3p]{elsarticle}

\usepackage[hidelinks]{hyperref}

\usepackage{libertinus,libertinust1math,courier}
\usepackage[T1]{fontenc}
\DeclareUrlCommand\path{\urlstyle{sf}}

\makeatletter
\let\@afterindenttrue\@afterindentfalse
\makeatother

\nopreprintlinetrue

\usepackage{array,booktabs,microtype,multirow,nameref,textgreek,xurl}

\biboptions{sort&compress}

\begin{document}

\begin{frontmatter}

\title{Digitization of Pathology Labs: A Review of Lessons Learned}

\author[mevisaddress]{Lars Ole Schwen\corref{correspondingauthor}}
\cortext[correspondingauthor]{Corresponding author}
\ead{ole.schwen@mevis.fraunhofer.de}

\author[chariteaddress]{Tim-Rasmus Kiehl}

\author[chariteaddress]{Rita Carvalho}

\author[chariteaddress]{Norman Zerbe}

\author[mevisaddress]{André Homeyer}

\address[mevisaddress]{Fraunhofer Institute for Digital Medicine MEVIS, Max-von-Laue-Str. 2, 28359 Bremen, Germany}
\address[chariteaddress]{Charité – Universitätsmedizin Berlin, corporate member of Freie Universität Berlin and Humboldt Universität zu Berlin, Institute of Pathology, Charitéplatz 1, 10117, Berlin, Germany}

\begin{abstract}
  Pathology laboratories are increasingly using digital workflows.
  This has the potential of increasing lab efficiency, but the digitization process also involves major challenges.
  Several reports have been published describing the individual experiences of specific laboratories with the digitization process.
  However, a comprehensive overview of the lessons learned is still lacking.

  We provide an overview of the lessons learned for different aspects of the digitization process, including digital case management, digital slide reading, and computer-aided slide reading.
  We also cover metrics used for monitoring performance and pitfalls and corresponding values observed in practice.

  The overview is intended to help pathologists, IT decision-makers, and administrators to benefit from the experiences of others and to implement the digitization process in an optimal way to make their own laboratory future-proof.
\end{abstract}


\end{frontmatter}

\section{Introduction}

\subsection{Background}

Digitization of pathology labs has been a growing trend for over a decade \cite{PanShaCar2018}.
This process can be coarsely categorized into three major steps: 1) digital case management, 2) digital slide reading, and 3) computer-aided slide reading.

Digital case management refers to digitizing workflows for managing specimens and slides, as well as reporting \cite{FraCapGug2021}.
This involves techniques such as barcoding cassettes and slides rather than labeling them by hand, composing reports digitally rather than on paper, and employing speech recognition for reporting \cite{HodCoi2015, AlACho2003, JohLapLon2014, PinBycTsu2023}.
In particular, the lab is equipped with a laboratory information system (LIS), sometimes referred to as laboratory information management system (LIMS) or specifically anatomical pathology LIS (AP-LIS), at this point \cite{ParPanSha2012, GuoBirFar2016, MatMar2017}.
This step involves techniques similar to those used when digitizing non-pathology labs \cite{HarMcD2008, YusAri2016, SepYou2013} and goes hand in hand with lab automation \cite{NauChu2019, BoyHaw2012}.
Many labs have already implemented this step, with key benefits of reducing tedious error-prone manual `specimen bookkeeping,' easier access to other patient data, and simplified billing.

Digital slide reading means pathologists view slides at computer workstations rather than under an analog microscope.
This requires acquisition of digital images of the slides with whole-slide scanners.
Using digital slide images obviates the need for transporting and handling physical slides afterward.
Viewer software includes support tools, such as distance measurement or digital annotations, and checks whether the entire slide has been viewed \cite{LPON2016}.
The main benefits are a more convenient case organization and distribution of work \cite{Bet2021, HoxAriPar2012, MaySmeSem2022}, allowing remote consultation \cite{LisBelBil2020, RetAneMor2020a} and the possibility to work from home \cite{AraAmaPer2021, ArdReuHam2021, HanReuArd2020, LisBelBil2020, LujSavSha2021, RamTejUth2021, RaoKumRaj2021, SamSroJot2020, SchGenYar2021}.
Moreover, cases at multidisciplinary meetings (tumor boards) can be presented more efficiently in digital form \cite{BerMorRin2020, LPON2016, ThoMolLun2014}, and teaching/training \cite{VaiMorPou2017, SagAri2018, SamSroJot2020}, as well as research \cite{BaxEdwMon2022} become more convenient.
Many labs around the world have recently taken or are planning this step.

Computer-aided slide reading involves advanced algorithmic image analysis tools for quantification, detection of hotspots/regions of interest, tissue classification, and more \cite{AefZarBuc2019, CuiZha2021, LaaLitCio2021, RakTosShi2020}.
Such analyses are run on demand or precomputed, possibly before a pathologist even looks at the case for the first time \cite{FlaFraSon2022, TemLopViv2022}.
Algorithmic results can be shown to the pathologist immediately or in a `second read' system only if they differ substantially from the human assessment \cite{PanQuiBie2020}.
Automated quantitative and large-scale assessment complements human diagnostic strengths.
This type of computational support is a growing field, and extensive research (see, e.g., \cite{EchRinBri2020, SerIonQur2019, SriCigMar2020, TizPan2018, BerSchRim2019}) has resulted in a small but increasing number of tools matured to approved diagnostic solutions \cite{ColPitOie2019, AcsRanHar2020, HomLotSch2021, MueDanVok2021, SteCheMer2021, BerMcCByc2023}.

\subsection{Related Work}

Feasibility of digital slide reading has been shown in numerous studies finding non-inferiority to classical light microscopy \cite{AzaMilKim2020, GoaRanWil2017, KusRifRis2022, SacRamRak2016, WilDaCGoa2017}.
A number of guideline publications for digitizing pathology labs with general recommendations \cite{CroFurIga2018, EvaBroBui2021, FraLImAme2021, Gar2016, HanArdReu2022, Lee2018a, LPON2016, WilBreAsl2020} are available.
Other guidelines focus on specific aspects, such as the clinical case \cite{WilBotTre2017} and business case \cite{LujQuiHar2021, WilBotCla2018} of digital pathology, on the scanning process \cite{ZarBowAef2019}, and training \cite{WilTre2019}.

Moreover, various labs have reported their experiences when transitioning to digital workflows \cite{AlJHuiNap2012, ArdReuHam2021, BabGunWal2021, Bai2015, ChaSadYao2022, CheAzhSng2016, ChoPalFis2019, EloValCur2021, EvgAleAnd2022, FerMonCoe2023, FraCapGug2021, FraGarZan2017, HagTolSch2021, HarPanMcH2017, LPON2016, LujSavSha2021, MonMonFra2022, RamTejUth2021, RaoKumRaj2021, RetAneMor2020a, SchGenYar2021, StaNguDie2020, StaNguSpo2019, StaVetHui2013, TemLopViv2022, ThoMolLun2014, VodAgh2018, VolChoEva2018, ZhaWuxDin2015}, see Table \ref{tab:experienceReports}.
Some labs are driven by enabling part of the staff (pathologists) to work from home \cite{LujSavSha2021, RamTejUth2021, RaoKumRaj2021, SchGenYar2021, ArdReuHam2021, AraAmaPer2021}.
A recent preprint \cite{PinBycTsu2023} surveyed 127 labs in Asia and Europe, of which 72 and 23 had established digital slide reading and computer-aided slide reading, respectively.
In addition to reports on the digitization process as a whole, others have focused on individual aspects.
Among them are the introduction \cite{GuoBirFar2016} and evaluation \cite{MatMar2017} of a LIS in a pathology lab and other clinical labs \cite{GosSinCha2010, LamJac2012}, the feasibility of multi-site reporting \cite{MaySmeSem2022}, the spent on lab logistics \cite{BaiBucLee2018}, pitfalls of whole-slide imaging \cite{AtaTosVer2021}, archive scanning \cite{HuiLooBri2010}, the impact on training \cite{BroColRit2019, BroWinCoo2022}, experience of pathologists involved \cite{LisBelBil2020, CapGibBel2022}, equivalency of reporting based on digital and analog workflows \cite{HanReuHam2019, AraAmaPer2021}, and an analysis of efficiency and costs \cite{HanReuSam2019}.

\begin{table}[t!]
  \centering
  \caption{Overview of publications reporting their experience with digitizing pathology laboratories.}\label{tab:experienceReports}

  \renewcommand*{\arraystretch}{1.2}\small
  \begin{tabular}{>{\raggedright}p{0.12\textwidth} >{\raggedright}p{0.12\textwidth} >{\raggedright}p{0.56\textwidth} >{\raggedright}p{0.09\textwidth}}
    \toprule
    Continent        & Country         & Institution                                                                                               & Literature                                                        \tabularnewline
    \midrule
    Asia             & India           & Strand Life Sciences - Health Care Global Cancer Hospital, Bengaluru (attached to hospital chain)         & \cite{RamTejUth2021}                                              \tabularnewline
                     &                 & Tata Memorial Hospital, Mumbai                                                                            & \cite{RaoKumRaj2021}                                              \tabularnewline
                     & Singapore       & Singapore General Hospital                                                                                & \cite{CheAzhSng2016}                                              \tabularnewline
    Europe           & Italy           & Cannizzaro Hospital, Catania                                                                              & \cite{FraGarZan2017}                                              \tabularnewline
                     &                 & Gravina Hospital in Caltagirone (with seven hospitals in Catania area)                                    & \cite{FraCapGug2021}                                              \tabularnewline
                     & Netherlands     & Laboratory for Pathology East Netherlands LabPON (working with different hospitals)                       & \cite{Bai2015, LPON2016}                                          \tabularnewline
                     &                 & University Medical Centre Utrecht                                                                         & \cite{AlJHuiNap2012, StaVetHui2013, StaNguSpo2019, StaNguDie2020} \tabularnewline
                     & Norway          & Førde Central Hospital                                                                                    & \cite{VodAgh2018}                                                 \tabularnewline
                     & Portugal        & Germano de Sousa - Centro de Diagnóstico Histopatológico (CEDAP), Coimbra (private lab)                   & \cite{FerMonCoe2023}                                              \tabularnewline
                     &                 & Institute of Molecular Pathology and Immunology of the University of Porto (IPATIMUP)                     & \cite{EloValCur2021}                                              \tabularnewline
                     & Spain           & Catalan Health Institute (8 hospitals)                                                                    & \cite{TemLopViv2022}                                              \tabularnewline
                     &                 & Granada University Hospital (2 central teaching and 2 peripheral hospitals)                               & \cite{RetAneMor2020a}                                             \tabularnewline
                     & Sweden          & Kalmar County Hospital and Linköping University hospital                                                  & \cite{ThoMolLun2014}                                              \tabularnewline
                     & Wales           & Glan Clwyd Hospital (hub)                                                                                 & \cite{BabGunWal2021}                                              \tabularnewline
    Europe/Africa    & Portugal et al. & IMP Diagnostics, Porto/Lisbon (working with Hospitals from Portugal, Angola, Cape Verde, and Mozambique)  & \cite{MonMonFra2022}                                              \tabularnewline
    Europe/Asia      & Russia          & UNIM Moscow (central lab serving all Rusia)                                                               & \cite{EvgAleAnd2022}                                              \tabularnewline
    North America    & Canada          & University Health Network and Laboratory Medicine partners, Ontario                                       & \cite{VolChoEva2018, EvaVajHen2020}                               \tabularnewline
                     & USA             & City of Hope National Medical Center, Duarte, CA                                                          & \cite{ChaSadYao2022}                                              \tabularnewline
                     &                 & University of California, Los Angeles (with 3 other University of California sites and 3 other hospitals) & \cite{ChoPalFis2019}                                              \tabularnewline
                     &                 & Mount Sinai Hospital, NY                                                                                  & \cite{HagTolSch2021}                                              \tabularnewline
                     &                 & Memorial Sloan Kettering Cancer Center, New York (NY)                                                     & \cite{ArdReuHam2021, SchGenYar2021}                               \tabularnewline
                     &                 & Ohio State University Comprehensive Cancer Center                                                         & \cite{LujSavSha2021}                                              \tabularnewline
                     &                 & \textgreater 20 hospitals across Western Pennsylvania                                                     & \cite{HarPanMcH2017}                                              \tabularnewline
    N. America/ Asia & USA+China       & University of Pittsburgh Medical Center and KingMed Diagnostics (China)                                   & \cite{ZhaWuxDin2015}                                              \tabularnewline
    \bottomrule
  \end{tabular}
\end{table}

\subsection{Contribution}

The experience reports cited above describe general experiences on different aspects of the digitization process.
They often also mention performance characteristics and pitfalls observed in the respective labs and contain data on the frequency of individual issues when digitizing a pathology lab to the point of digital slide reading.
In this article, we provide a systematic overview of the lessons learned with regard to key aspects in the implementation and operation of digital pathology workflows.
We also cover the quantitative assessment of these aspects and summarize data reported by the various labs.
While many reports only cover the digitization process to the point of digital slide reading, we also provide a perspective for the next step of using image analysis algorithms for computer-aided slide reading.

\section{Methods: Literature Search}

This overview is based on an exploratory literature search based on articles with titles deemed relevant in the \href{https://www.sciencedirect.com/journal/journal-of-pathology-informatics/}{Journal of Pathology Informatics} since 2020 and the 2022 special issue \href{https://www.mdpi.com/journal/diagnostics/special\_issues/Digital\_Pathology\_Successful\_Implementations}{Digital Pathology: Records of Successful Implementations} of MDPI Diagnostics.
From this initial collection, we iteratively included papers cited therein that fell within the same scope and searched for related publications using related keywords.

\section{Results}

\begin{figure}[t!]
  \centering
  \includegraphics[width=\linewidth]{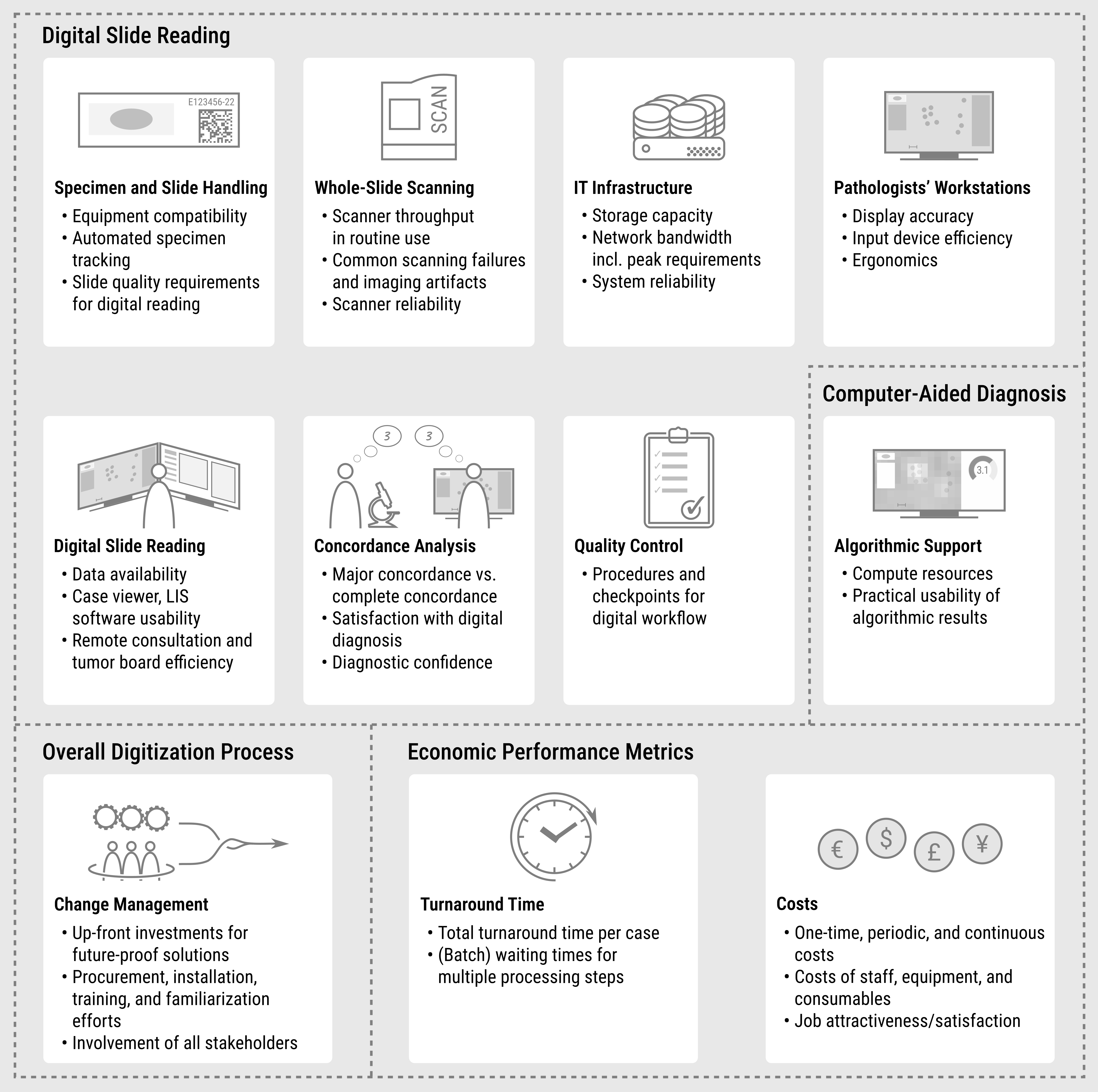}

  \caption{Most relevant aspects of implementing and assessing the digitization process in pathology labs addressed in this review.}\label{fig:overview}
\end{figure}

The lessons learned in this section are structured in a bottom-up fashion, see Figure \ref{fig:overview} We first consider the transition of a pathology lab to digital slide reading (including digital case management), followed by additional aspects applicable when introducing computational support by image analysis algorithms.
We finally take a more general perspective on digitization, including the assessment by macroscopic economic metrics.

\subsection{Digital Slide Reading}\label{sec:digitalSlideReading}

Switching from a diagnostic workflow using analog microscopes to digital slide reading implies many changes throughout the entire workflow.
The aspects related to these changes are structured here according to the path that a specimen and its scans take through the lab.

\subsubsection{Specimen and Slide Handling}\label{sec:SpecimenHandling}

Scanning and reading digital slides changes the (quality) requirements for specimen preparation.
This may require using other hardware, e.g., to avoid incompatibility of equipment between staining and scanning racks \cite{FraGarZan2017}, or other chemicals/consumables to facilitate subsequent scanning \cite{StaNguSpo2019}.
Moreover, equipment may need to be relocated to avoid delays in digital workflows \cite{FraGarZan2017}.
These changes lead to changes in effort (i.e., personnel cost), cost of consumables, and processing times per slide.
Moreover, efforts and costs per case may be lowered by reduced requests for additional staining, e.g., fewer immunohistochemistry (IHC) stain requests \cite{HanReuSam2019}, and by automated specimen tracking by scanning barcodes instead of manually transcribing identifiers \cite{FraGarZan2017}.

\paragraph{Pitfalls}

Various specific issues with slide preparation were reported in the literature.
Barcodes from external labs can lead to confusion \cite{HanReuSam2019}; this study suggests covering such barcodes with erasable chalk ink instead of stickers to avoid covering relevant label information and cumbersome removal of stickers.
The lab's own barcodes may be printed incorrectly (in 0.5\% of cases, this resulted in reading errors in \cite{FraGarZan2017}).
Accessioning errors may occur: one study reported a reduction from 6.3\% to 0.5\% when switching to a digital workflow \cite{FraCapGug2021}.
Slides may be sectioned too thickly, leading to problems with imaging/focus when scanning \cite{AraAmaPer2021, FraGarZan2017, AlJHuiNap2012}; tissue folding causes similar problems \cite{AlJHuiNap2012}.

The final slides may contain excess glue \cite{FraGarZan2017}, air bubbles \cite{FraGarZan2017} or frothy medium under the coverslip \cite{PatShaErc2022}, tissue extending beyond the coverslip \cite{PatShaErc2022}, dirt, dust, markings, mounting media \cite{PatShaErc2022}, or other artifacts \cite{AraAmaPer2021}.
Only some of these issues can be resolved by cleaning or keeping slides clean with the associated efforts \cite{ThoMolLun2014}.

Many of these issues require the production of a replacement slide.
It is possible to measure the frequency of such issues as a percentage of slides or cases affected.
Reported data about such issues is scarce, as most sources quantify them as scanning issues.
The issues reported in \cite{FraGarZan2017} amounted to around 1\% of slides (without breakdown in individual issues).

\subsubsection{Whole-Slide Scanning}\label{sec:WholeSlideScanning}

The number and models \cite{SchGenYar2021, Tri2022} of whole-slide scanners in a pathology lab need to meet the requirements \cite{NamChoJun2020} for efficient operation.
Various metrics describing the capacity and availability of scanners can help in procurement decisions and in determining the need for change.

\paragraph{Scanner Capacity}

Scanning capacity needs to be sufficient for the number of slides to be scanned per working day/month/year.
Reported numbers of slides per year per scanner range between 40k and 148k slides per scanner and year \cite{MonMonFra2022, FlaFraSon2022, TemLopViv2022, HoxAhlStr2014, SchGenYar2021, LPON2016, StaNguSpo2019, LujSavSha2021, ArdReuHam2021, HuiLooBri2010, ArdKleMan2023}.
These numbers are, however, difficult to compare, as they depend on usage (e.g., whether there are fluorescence scans), scanner models (capacity, scanning speed), and specific requirements (e.g., \cite{SchGenYar2021} mentions 52k routine vs. 93k archive scans per scanner and year).
Scanning capacity does not only need to support the average demand but also peak times (e.g., \cite{StaNguSpo2019} found a requirement of 800 slides per day, but 120 per hour during peak times; \cite{EvgAleAnd2022} report an average number of slides per day of 699 with a maximum of 1210, i.e., 173\% of the average).
Moreover, the volume of unattended (overnight) batch scanning \cite{Lee2018a, PatBalChe2021, StaNguDie2020, FraCapGug2021, MonMonFra2022, RamTejUth2021, RetAneMor2020a, TemLopViv2022, ThoMolLun2014} should be considered.

\paragraph{Scanner Availability}

In case of scanner downtime, reserve scanning capacity may be needed.
Observed scanner downtime should be compared to required or specified uptime, e.g., 99.9\% \cite{StaNguSpo2019}.
Similarly, scanner throughput in routine practice differs from advertised throughput \cite{GenRemUnt2021, ArdKleMan2023}, e.g., due to scanner parameters in routine (such as handling operations) being different from the specification \cite{MatSch2022}.
Scan times per slide \cite{EloValCur2021, RetAneMor2020a, RamTejUth2021, RaoKumRaj2021, HanReuArd2020, FraGarZan2017, EloValCur2021, LPON2016, HanReuHam2019, AlJHuiNap2012, FerMonCoe2023} depend on tissue area per slide and scanning resolution.
For this assessment, the time of handling/loading requiring personnel efforts and the actual scanning (automated) should be distinguished.
There is potential for speed-up by optimizing slides and scanner parameters: \cite{MonMonFra2022} reported a median of 98 s per slide in the test phase, which they could reduce to a median of 74 s after optimizing fragment placement.

For maintenance and repairs, costs of external personnel and replacement parts for repairs should be considered (possibly part of a lease or service/maintenance contract).
In addition to scheduled maintenance, unplanned scanner downtimes \cite{MaySmeSem2022} may occur.
These are quantifiable by an overall uptime percentage \cite{StaNguSpo2019}, frequency of downtimes or the number of shifts affected, and downtime duration.
One study reports a range of 28 to 85 minutes for downtime durations \cite{HarPanMcH2017}.
Another study reported between 69 and 106k successful scans per service call \cite{ArdKleMan2023}.

\paragraph{Pitfalls}

Numerous specific pitfalls may occur during scanner operation.
These can be quantified as a percentage of slides affected or as a number of occurrences per shift/day/month.
Scan failure/re-scan rates are generally required and observed to be in the range below 2\% \cite{FerMonCoe2023, MonMonFra2022, AtaTosVer2021, SchGenYar2021, LujSavSha2021, HanReuArd2020, FraGarZan2017, RaoKumRaj2021, PatShaErc2022} or below 1\% \cite{FraCapGug2021, RetAneMor2020a, RamTejUth2021, LisBelBil2020}, with studies reporting as much as 7\% \cite{HanReuHam2019}, an increase from 84\% initial to 98\% final scan success rate \cite{MonZarAre2022}, and only 0.5 to 4.5 scan errors per month \cite{HarPanMcH2017}.
Scanners can, in principle, report some types of failures: slide being stuck to the prongs of the robotic arm (0.7\% of slides reported in \cite{RetAneMor2020a}); barcode detection failure (0.5 to 1.2\% \cite{HanReuArd2020, RetAneMor2020a, FraGarZan2017, MonMonFra2022, VolChoEva2018}); scanner breaking glass slide \cite{ThoMolLun2014, FraGarZan2017}; slides sticking to scanner racks if not dried properly \cite{FraGarZan2017, StaNguSpo2019}; obstruction of slide labels \cite{FraGarZan2017}; camera error \cite{HarPanMcH2017}; robot error (slide sideways) \cite{HarPanMcH2017}; robot dropping slide (4.5 times per month \cite{HarPanMcH2017}), e.g., as a consequence of label or cover slide overhanging, leading to grip failure and dropped slide, possibly resulting in broken slide \cite{HarPanMcH2017}, slide dropped (0.07\% \cite{PatShaErc2022}); or unexpected slide in gripper \cite{HarPanMcH2017}.
Other issues can only be detected by systematic post-scan image QC \cite{MonMonFra2022} or by the pathologist reading the image.
Such issues include incomplete images (not all tissue visible in the image) \cite{AlJHuiNap2012, CapGibBel2022, JahPlaMoi2020, PatShaErc2022, VolChoEva2018}, also denoted as tissue dropout and reported as occurring in 29.5\% of the slides \cite{MutMilHwa2022} or 19\% of the slides with 13\% of tissue on average \cite{AtaTosVer2021}.
However, adjusting scanner settings can drastically reduce these numbers: among more than 2.2 million scanned slides reported in \cite{PatShaErc2022}, tissue was skipped in only 0.27\%.
Another issue is blurry images due to focus errors, reported in 0.5 to 3.5\% of slides \cite{FraGarZan2017, MonMonFra2022}, necessitating re-scans in 0.5\% of the slides in a different study \cite{VodAgh2018}.
Focus errors may be due to confusing automatic focus \cite{HarPanMcH2017}, in which case cleaning the slide can help.
Similarly, the objective lens may require cleaning (for which the scanner needs to be taken offline), resulting in `Venetian blinding' artifacts \cite{PatShaErc2022}.
Image stitching may fail \cite{PatShaErc2022}, in which case scanner support is needed.

Besides scans clearly failing, issues can also result in imaging artifacts that may or may not affect the diagnosis \cite{MonMonFra2022, RetAneMor2020a, ClaDohRan2022}, e.g., chromatic aberrations \cite{PinBycTsu2023}.
Here, percentages of affected cases can be useful metrics, requiring feedback from a later step in the digital workflow.
When running unsupervised overnight batch scans, such failures can go unnoticed and cause delays until images can be re-scanned after resolving the issues on the next working day \cite{FraGarZan2017, StaNguDie2020, FraLImAme2021}.
Continuous monitoring can mitigate this problem and, if necessary, alert a technician on call \cite{TemLopViv2022}, which implies additional costs for this position.

Data on scanner issues need to be communicated appropriately to avoid an incorrect perception of the frequency of scanner issues \cite{SmiJohZeu2022}.
In labs with more than one scanner, both the frequency of issues and characteristics of the resulting images may differ between scanners \cite{MutMilHwa2022}.
Also, it is possible to quantify this issue via the percentage of diagnoses affected.
Scanner calibration and alignment of scanner settings can mitigate this issue.
In particular, color calibration is needed along the entire optical pipeline.
Scanners can be calibrated \cite{HuiLooBri2010, ThoMolLun2014, JahPlaMoi2020} using slides with defined color patches \cite{GriTre2017, ClaTre2016, BauHasYag2014}.
Details on the calibration of displays and the impact of a loss of calibration are addressed below.

\subsubsection{IT Infrastructure}\label{sec:ITInfrastructure}

\paragraph{Storage Capacity}

The key challenge in designing IT infrastructure for digitized pathology labs is the very large amount of image data compared to other disciplines.
It is typically on the order of 100s of MB to several GB per slide \cite{RaoKumRaj2021, HanReuArd2020, FraGarZan2017, EloValCur2021, SchGenYar2021, HanReuHam2019, MonMonFra2022, RetAneMor2020a, ThoMolLun2014}.
Image size depends on the size of the tissue, imaging resolution and modality (e.g., fluorescence), and image compression settings.
These amounts of data impact both storage capacity and network infrastructure for transferring data.
Calculating required storage per slide, the typical tissue size per slide (small biopsy vs. large specimens \cite{EloValCur2021}), scanning resolution \cite{SchGenYar2021}, imaging modality (light vs. fluorescence microscopy \cite{EloValCur2021}), image quality (lossy compression) settings \cite{ZarBowAef2019, Gar2016, StaVetHui2013, CroFurIga2018}.
While scanning resolution and compression settings can be optimized as a compromise between image quality and file size, further potential for optimization of fragment placement on slides for optimal background omission of 26\% has been reported \cite{MonMonFra2022}.

The total storage space required depends on how many slides are scanned per day (on average) and how long digital data is retained.
A common strategy is to implement multi-tier storage, e.g., store `current' cases on hard disks for fast access and archive data on slower but cheaper tapes \cite{LujQuiHar2021, ArdKleMan2023, PinBycTsu2023}.
Already in 2010, one study reported 6 TB disk and 120 TB tape storage \cite{HuiLooBri2010}.
A more recent study reported a setup with 20 TB storage for collecting data from scanners, 360 TB for storing cases of one year for immediate availability, and 1 PB tape archive for cases from three years \cite{RetAneMor2020a}.
Another study describes plans for 150 TB to store cases from three months and multiple PB for long-term storage in the clinical picture archiving and communication system (PACS) \cite{WilBozKin2022}.
A different approach is to archive only part of the image data, e.g., one \cite{Gau2022} or a few \cite{DasJonMer2021} slides per sample, resulting in smaller long-term storage requirements, e.g., 220 TB for ten weeks of current cases plus 25 TB per year \cite{Gau2022}.
The cost for fast-access and long-term storage was reported to be 150 and 60 € per TB and year, respectively \cite{Gau2022}.
Storage capacities should match requirements for how long cases need to be accessible quickly; how long digital data shall be archived; how long after scanning slides are typically accessed \cite{SchGenYar2021}; and how quick the retrieval from the archive is supposed to be \cite{HanReuSam2019, LujQuiHar2021} (e.g., retrieving on demand vs. pre-fetching when the patient is known while a new case is being processed in the lab).
While archiving image data on cheaper but slower storage is efficient, metadata for searching is small and can be kept in fast storage \cite{LujSavSha2021}.
Total storage was reported on the order of 10s of TB to several PB \cite{HuiLooBri2010, SchGenYar2021, RetAneMor2020a, FraCapGug2021, EloValCur2021, TemLopViv2022, LPON2016, ChaSadYao2022, AlJHuiNap2012, LujSavSha2021}.
Depending on how storage is implemented (as a storage solution/image management system/vendor-neutral archive specifically for pathology or as part of an institutional PACS \cite{JanLinStr2023}), deviations between the expected amounts of data and those eventually observed need to be addressed at different locations.

\paragraph{Network Bandwidth}

Data transfer occurs mainly from scanners to image storage, from storage to workstations, between sites, and possibly to off-site backups.
Network bandwidth for local core infrastructure is commonly in the range of 100 Mbit/s to 10  Gbit/s \cite{FerMonCoe2023, FraCapGug2021, MonMonFra2022, TemLopViv2022, SchGenYar2021, RamTejUth2021, FraGarZan2017, FraCapGug2021, EloValCur2021, TemLopViv2022, LujSavSha2021, ArdKleMan2023} and should be designed to handle the load at peak times.
For connecting workstations, a recent study reported that 300 Mbit/s would have been sufficient \cite{TemLopViv2022}, so standard gigabit ethernet should suffice.
Only single cases need to be transferred with low latency when working remotely.
Speeds between 4 and 500 Mbit/s were reported for pathologists working from home \cite{RaoKumRaj2021, HanReuArd2020, RamTejUth2021}.
Faster connections are needed for transmitting multiple cases between sites \cite{ChoPalFis2019}, and in case of off-site backups, e.g., if 6 TB of data per day need to be transferred \cite{TemLopViv2022}, this corresponds to a continuous (24/7) transfer rate of 556 Mbit/s.
A relevant issue when working remotely are slow internet connections, particularly in rural areas \cite{MaySmeSem2022, LujSavSha2021}.
One study reported transfer times of 2.5 to 28 minutes for a single case \cite{ChoPalFis2019}, which requires planning of data transfer to minimize delays in data usage.
Besides the remote connections, central VPN access can become slow during concurrent access \cite{MaySmeSem2022} and should be designed to accommodate peak load.

\paragraph{Pitfalls}

Standard issues in IT infrastructure, such as network interruption or failure of storage media, also affect digitized pathology labs.
Network downtimes are reported as rare (once since installation \cite{BaiBucLee2018}; one `system error' in 6606 cases \cite{CheAzhSng2016}).
It is possible to mitigate these issues with standard IT measures, such as using a redundant array of independent disks (RAID) for the possibility of replacing failing storage media without data loss if such precautions are not already part of operating a hospital PACS.
Given the cost of storing these amounts of data and depending on whether glass slides are archived as well, having no offsite backup or no backup at all may be valid strategies \cite{HuiLooBri2010, RetAneMor2020a, StaNguDie2020}.
Hence, if data is lost, efforts for data recovery may range from none (accept loss of image data) to substantial (re-scan), albeit for different persons.

With multiple software components needed in a digitized pathology lab, interoperability between systems in the initial installation and after updates needs to be implemented and tested \cite{Gau2022}, which is largely simplified if software components follow the applicable standards.
For labs part of a hospital, this also involves the connection to and interoperability with hospital IT.
This requires personnel efforts by the IT team and, to some extent, also by persons in the lab and may cause implementation delays.

\subsubsection{Pathologists' Workstations}\label{sec:PathologistsWorkstations}

\paragraph{Displays}

In this section, `workstation' refers to the hardware setup; slide viewer software is a topic in the next section.
Pathologists' workstations in a digitized lab are similar to office-grade desktop setups with a PC.
They usually have two displays \cite{PinBycTsu2023}, sufficiently powerful graphics cards, and dedicated input devices.
Displays should be sufficiently large and have sufficiently high resolution to provide enough detail \cite{Gar2016, RojBue2015, RanAmbMel2014} but not too large to require excessive head movement \cite{LPON2016}.
Typically reported sizes range from 24 to 32 inches (diagonal) \cite{FerMonCoe2023, RetAneMor2020a, HanReuHam2019, StaNguSpo2019, FraCapGug2021, EloValCur2021, TemLopViv2022, HanReuArd2020, ThoMolLun2014, RamTejUth2021, LisBelBil2020, BabGunWal2021, FraGarZan2017, LPON2016, PinBycTsu2023} with image refresh rates of typically 60 or 75 Hz \cite{BabGunWal2021, FraCapGug2021, FerMonCoe2023} and resolution from 2 to 8 MPx \cite{FerMonCoe2023, RetAneMor2020a, HanReuHam2019, StaNguSpo2019, FraCapGug2021, EloValCur2021, TemLopViv2022, HanReuArd2020, ThoMolLun2014, RamTejUth2021, LisBelBil2020, BabGunWal2021, FraGarZan2017, RaoKumRaj2021, PinBycTsu2023}.
To compare displayed image resolution to analog microscopes, the entire optical pipeline, including digital zoom, must be considered \cite{RojBue2015, RanAmbMel2014, SelFilHof2013}.
The image resolution of the acquired whole-slide image depends on the objective magnification (e.g., 40×), subsequent optics, and the image sensor, leading to image resolutions of typically hundreds of nanometers per pixel \cite{SelFilHof2013}.
The amount and detail of information visible on displays depend on display size and resolution, typically a few hundred pixels per inch for desktop monitors and lower for large displays or projected images \cite{RojBue2015, RanAmbMel2014}.
Incorrectly treating resolution (e.g., assuming that all images acquired with objective magnification have the same image resolution) may result in incorrect size assessments.

A dual-monitor setup was found beneficial \cite{ThoMolLun2014, RetAneMor2020a, HanReuHam2019} with, e.g., one display used for viewing images and one (smaller, not necessarily medical-grade) display for LIS data, patient records, and more.
A third monitor can be used for other applications, e.g., e-mail or other communication \cite{Gau2022}.
When using office- or consumer-grade hardware \cite{JahPlaMoi2020, PinBycTsu2023}, color depth is usually 24 bit (8 bit per red/green/blue channel) \cite{BabGunWal2021}, and color calibration is necessary \cite{BauHasYag2014, ShrHul2014, KruSilHas2012, AbeOuiWil2020, ClaRevBre2016, ClaTre2016, WriClaDun2020}.
Medical-grade displays offer a higher color depth of 10 or 12 bits per channel \cite{AbeOuiWil2020} and auto-calibration \cite{RetAneMor2020a} for displaying images more accurately, as well as higher refresh rates \cite{AbeOuiWil2020} to reduce eye strain.
Lack or loss of calibration due to device aging \cite{AvaEspXth2016} decreases the perceived contrast of clinically relevant features \cite{ZarBowAef2019, KimRosAva2014}, which may be quantified via a percentage of cases affected.
This issue can be mitigated by regular calibration \cite{AvaEspXth2016, WriClaDun2020, ZarBowAef2019, JahPlaMoi2020, RetAneMor2020a}.
In addition to single-user workstation displays, larger displays or projectors may be used, e.g., for teaching and interdisciplinary conferences \cite{TemLopViv2022}.
The main pitfalls regarding displays are inter-display differences, in particular, hard-to-calibrate projectors or large displays used in tumor conferences or cell phones/tablets used to view image data \cite{ZarBowAef2019, ClaMunWil2020}.

\paragraph{Computers and Input Devices}

Memory (RAM) was reported to be 3 to 32 GB \cite{FerMonCoe2023, RaoKumRaj2021, BabGunWal2021, FraGarZan2017, FraCapGug2021, EloValCur2021, TemLopViv2022, RetAneMor2020a, LPON2016, HanReuHam2019, LujSavSha2021}.
Standard office equipment like mouse and keyboard are often complemented by specialized devices (such as game mice, 3D mice, trackballs, specialized pathology devices, trackpads and touch displays) for navigating slides more efficiently and to avoid ergonomic problems such as, e.g., hand/wrist pain \cite{MolLunFje2015, AlcCabAlb2016, StaNguSpo2019, WilIsmCha2020, SchGenYar2021, AlcTurNie2020, ClaDohRan2022, KruKal2007, LPON2016, BabGunWal2021}.
In addition, computer workstations pose other ergonomic challenges than microscope workstations \cite{WilBotCla2018, JahPlaMoi2020, SmiJohZeu2022}, such as the danger of back/neck pain \cite{BabGunWal2021} or eye strain \cite{KruKal2007, LPON2016, StaNguSpo2019}, cf. \cite{KruKal2007} for similar observations in radiology.
One advantage of working digitally is that pathologists do not need to handle physical slides with residual chemicals \cite{SmiJohZeu2022}.
For a quantitative comparison, the number of reports of ergonomic problems per person and time can be considered, ideally assessed by surveys \cite{ThoMolLun2014, StaNguSpo2019, JahPlaMoi2020} and mitigated proactively rather than by waiting for pathologists to call in sick.
For assessing optimizing individual efficiency and job satisfaction, users' personal preferences regarding the hardware setup should be taken into account \cite{RanAmbMel2014, ClaMunWil2020, WilBotTre2017}.

\paragraph{Pitfalls}

When working from home, a different technical setup than in the lab (in particular, the display \cite{AraAmaPer2021}) may mean that cases take longer or diagnoses may differ \cite{RaoKumRaj2021, LisBelBil2020}.
Moreover, clinical and other data inaccessible from remote \cite{AraAmaPer2021} or tools such as speech recognition may be unavailable \cite{AlcTurNie2020}, resulting in loss of efficiency (time needed per case).
Other pitfalls include basic computer connection mistakes, such as inadequate adapters leading to unnecessarily low display resolution \cite{AbeOuiWil2020} or PCs too small to fit specific graphics cards physically.
These issues can be mitigated using the same device setup at every work location, by providing secure remote access to all required data and tools, and by proper IT setup and maintenance.

\subsubsection{Digital Slide Reading}\label{sec:DigitalSlideReading}

Pathologists spend the majority of their work time viewing slides and reporting (61.9\% according to \cite{RanRudTre2015}, 36\% plus 34.6\% according to \cite{HoxAhlStr2014}), with typically 5 minutes spent per case \cite{RaoKumRaj2021, ClaDohRan2022, BorGlaWal2020, ThrWimSch2015}.
However, the time per case strongly depends on the specific diagnostic tasks and the difficulty of the case.
Even though this is only a small portion of the total turnaround time, slide reading is a bottleneck due to the required expertise.
Data on time savings by digital slide reading are equivocal \cite{GoaRanWil2017}, with studies reporting slow-down \cite{VelJukHox2008, ThrWimSch2015, MilGraHor2018, HanReuHam2019, BorGlaWal2020}, no change \cite{RanRudTho2014}, and speed-up \cite{Bai2015, VodAgh2018, ClaDohRan2022, BaiKhaLaa2023}.
A breakdown by subspecialty categories \cite{BaiKhaLaa2023} showed differences in reading times and in changes thereof between analog and digital slide reading.
Time spent in multidisciplinary meetings (tumor boards) has the potential to save time for multiple participants \cite{BerMorRin2020, LPON2016}.
Here, time savings are reported: 15 cases could be discussed in 60 minutes with a digital setup instead of 90 minutes, and 25 instead of 15 cases could be discussed in 90 minutes \cite{ThoMolLun2014}.

\paragraph{Benefits}

In case access to glass slides is possible (e.g., when not working from home), the need/preference to view glass slides \cite{LisBelBil2020, MaySmeSem2022}, or conversely, a digital adoption rate \cite{HagTolSch2021, ThoMolLun2014, HoxAhlStr2014}, can be quantified as a percentage of slides or cases.
Older studies from 2006--2013 report low digital adoption, with 17.9\% preference for glass \cite{AlJHuiNap2012} or only 38\% of cases signed out digitally \cite{ThoMolLun2014}.
More recent studies, probably also with higher expectations of digital adoption, report glass preferences below 10\% \cite{EloValCur2021, LisBelBil2020, RaoKumRaj2021} and even below 1\% \cite{VodAgh2018} with a decreased demand for glass after adopting and getting used to a digital workflow \cite{PatBalChe2021}.
These findings match several studies reporting that pathologists were skeptical in advance but happy with diagnosing digitally later \cite{BaiBucLee2018, CapGibBel2022, NamChoJun2021}.

Potential benefits are that digital case records allow for a better overview of the current workload for planning days efficiently \cite{Bet2021, HoxAriPar2012}; the possibility of workload leveling also between sites to achieve an even distribution of overtime \cite{HoxAhlStr2014, Bet2021, MaySmeSem2022}; the possibility to make individual workload transparent more easily \cite{BlaDawHam2017}; and the possibility of working from home \cite{LujSavSha2021, RamTejUth2021, RaoKumRaj2021, SchGenYar2021, ArdReuHam2021, AraAmaPer2021, HanReuArd2020, LisBelBil2020} at least for pathologists.
Moreover, a digital setup allows for automatic tracking of whether all tissue on the slide was at least displayed, making it harder to miss relevant tissue parts \cite{LPON2016}; easier consultation with (e.g., more experienced) colleagues, in-house and remotely without the danger of losing physical slides \cite{SmiJohZeu2022}, making cross-site `curbside consultations' possible \cite{LisBelBil2020, RetAneMor2020a}; easier interaction with other external resources \cite{Bet2021, HoxAriPar2012}; faster retrieval of archived slides (within minutes rather than hours \cite{RetAneMor2020a}).
This also allows for ad-hoc discussions with clinical colleagues \cite{BaiBucLee2018} and thus maintaining better relationships with clinicians/external collaborating colleagues \cite{Bet2021, HoxAriPar2012, LujSavSha2021}.

\paragraph{Pitfalls}

Pitfalls include that scans are not found \cite{AlJHuiNap2012} (which can also happen with glass slides); usability aspects, such as configurability of viewers/digital pathology system \cite{Bet2021, HoxAriPar2012} or the lack of unified user experience \cite{ThoMolLun2014}; and that digital slide reading may become over-confident on low resolution, missing detail \cite{AraAmaPer2021}.

Many of these aspects can be quantified as part of the time spent per case and/or as the percentage of cases in which the diagnosis was affected.
Digitization may save time, but easy availability might also increase the workload because second opinions can easily be requested and because the easy availability of prior slides might lead to an obligation to check them \cite{JahPlaMoi2020}.
Besides the time spent per case, other, opinion-based aspects of pathologists' work are affected by digitization.
These refer mostly to the perceived efficiency and quality of working and, ultimately, job satisfaction \cite{MaySmeSem2022, PatBalChe2021, HanReuSam2019, HanReuArd2020, RetAneMor2020b, HanReuHam2019, DroMilVos2022, KelSolGar2020, GarKunKel2020}, which is hard to quantify, but still important in a competitive job market.
Satisfaction with specific tools, job satisfaction in general, and potential for improvement can be assessed by surveys (rather than learning about such issues only via rates of persons quitting their jobs and in exit interviews \cite{Ham2007}).
Such surveys are typically qualitative with multiple choice and free text questions; questions with responses on Likert scales \cite{AlcTurNie2020, HanReuSam2019, JahPlaMoi2020, MatMar2017, PehHasCal2017} can yield semi-quantitative results.

\subsubsection{Concordance Analysis}\label{sec:ConcordanceAnalysis}

When all steps of the digital workflow are implemented in a lab, it is necessary to verify that the diagnostic results are non-inferior to the previous, analog workflow.
Studies investigating concordance typically focus on major discordance or the clinical impact (harm to patients) of discordant diagnoses \cite{WilDaCGoa2017, BabGunWal2021, AzaMilKim2020} and also report the extent of complete concordance.
Concordance between analog and digital diagnosis is measured in a metric suitable for the diagnostic task at hand and compared to inter-observer variability \cite{ThrWimSch2015, GoaRanWil2017, RaoKumRaj2021, WacDroTre2016, WacDroTre2016}.
Typically, the same pathologists view the cases for analog and digital assessment; hence many studies used a suitable washout period between the two assessments.
Two weeks are recommended \cite{EvaBroBui2021}, however, studies range from no washout period used \cite{ThoMolLun2014, LeeJedPan2018} to periods of more than one year \cite{AzaMilKim2020}: two weeks \cite{WilIsmCha2020, ChaSadYao2022, WilTre2019, RaoKumRaj2021, WacDroTre2016, CheAzhSng2016}, three weeks \cite{JukDroMar2011, ThrWimSch2015}, one month \cite{BabGunWal2021, AraAmaPer2021}, six weeks \cite{MilGraHor2018}, three months \cite{RamTejUth2021} and half a year \cite{VodAgh2018, HanReuHam2019}.
In contrast, reexamining immediately is more suitable when focusing on assessing differences \cite{LeeJedPan2018} or becoming familiar with differences in appearance (e.g., to build confidence) \cite{WilTre2019}.

Reported concordance for analog vs. digital diagnosis by pathologists is generally high \cite{AzaMilKim2020, GoaRanWil2017, KusRifRis2022, SacRamRak2016, WilDaCGoa2017, HanReuHam2019, MilGraHor2018}.
However, it varies between organs and use cases \cite{SacRamRak2016, BorGlaWal2020}, so appropriate literature values should be chosen to serve as a baseline for validating the digital workflow in a lab.
In addition to the concordance, diagnostic confidence/certainty per case \cite{FonKreNag2012, RaoKumRaj2021, JukDroMar2011} and satisfaction with digital diagnosis per case \cite{RamTejUth2021} can be recorded.
Confidence is typically measured using Likert scales, e.g., from 1 to 3 \cite{RaoKumRaj2021, FonKreNag2012}, 1 to 5 \cite{JukDroMar2011, PehHasCal2017}, 1 to 6 \cite{KruSilHas2012}, or 1 to 7 \cite{WilTre2019, RanAmbMel2014, CroFurIga2018, WilIsmCha2020, FerMonCoe2023} No substantial differences in diagnostic confidence were reported (6.7/7 to 6.9/7 digital vs. 7/7 glass \cite{WilIsmCha2020}; 4.0/5 digital vs. 4.1/5 glass \cite{JukDroMar2011}).

\subsubsection{Quality Control}\label{sec:QC}

In a digital workflow containing scanning as an additional step, suitable quality control (QC) procedures are need\-ed \cite{RamTejUth2021}.
Besides differences in quality requirements, the key considerations are to what extent issues are detectable automatically or whether human assessment is needed and when issues can be detected and resolved \cite{PatShaErc2022, HanArdReu2022}, ideally as early as possible.
Most effort is saved if foreseeable issues are already detected before slide scanning.
However, only some issues are visible to the naked eye and are already detectable at this point.
Certain handling (e.g., slide dropped) and imaging issues (e.g., blurred images) can be detected automatically and flagged for manual intervention.
After scanning each slide, a dedicated image QC step \cite{EloValCur2021} is a way to visually assess image completeness and quality to trigger re-scans immediately.
Issues detected by pathologists during diagnostic review need to be communicated back to earlier points in the workflow and cause more delays.
However, in case of low scan failure rates, re-scanning on pathologists' demand may be more cost-efficient than routinely checking all images \cite{FerMonCoe2023}.
In addition, dedicated quality assurance procedures can monitor slide and scan quality, e.g., by inspecting 1.5\% of daily scans by senior staff \cite{PatShaErc2022}.
Here, the number of QC steps and efforts for the QC procedures can be quantified, as well as the rate of issues caught in the different QC steps.
Efforts for additional (early) QC steps can be compared to efforts for recutting/restaining/rescanning slides if issues are detected unnecessarily late in the workflow.

\subsection{Computer-Aided Slide Reading}\label{sec:ComputerAidedSlideReading}

Implementing advanced algorithmic support for diagnosis comes with additional requirements.
The algorithms are additional software that needs to be licensed and integrated in the existing infrastructure, i.e., further interoperability requirements need to be met.
Even though regulatory approval of algorithmic solutions is based on analytic and clinical performance data, local validation and suitably adapted QC procedures are necessary \cite{LunLin2022}.

\subsubsection{Hardware Requirements}

Additional computational servers are needed and require sufficiently fast data transfer from and to data storage.
One study reports seven servers with 8 Tesla V100 GPUs each; 6 servers with 12 GTX 1080Ti GPUs, 12 Titan X GPUs, and 512 GB RAM each; 5 PB of global parallel file system (GPFS) high-performance storage for raw data; and 100 Gb/s network backbone connection internally and 1 to 10 Gb/s externally \cite{SchGenYar2021}.
Requirements clearly depend on the volume of slides to be processed and the computational demands of the algorithms to be used in the foreseeable future.
Analyses can be run on demand (in manually defined regions of interest defined or the entire whole-slide image) or as automatic preprocessing of whole-slide images prior to slide reading.
In both cases, peak computational workload should not exceed computational resources to avoid a bottleneck and, as a consequence, delays.
Compute resources can be provided on premises or in the cloud \cite{JanLinStr2023}.
Cloud resources have the potential of easier scaling to peak load and compensating for outages but require sufficient data transfer and privacy considerations.

\subsubsection{Benefits}

Using advanced algorithmic support for diagnosis has the potential to reduce the reading time per case and to reduce the need for second opinions.
Additional potential of automatic preprocessing is in better prioritization \cite{MayGooMen2023} and assignment of cases, e.g., sending cases that are likely difficult directly to specialists.
Algorithmic support might also reduce the need for IHC stains if more information is extracted from H\&E stains.
Conversely, algorithmic preprocessing might also detect the need for additional IHC stainings earlier, leading to a faster feedback loop.
In contrast, algorithmic analyses started on demand may cause delays, frustrating users.
While algorithmic processing time does not necessarily have a big impact on the total turnaround time, it may reduce pathologists' efforts for part of the cases and leave them more time for more difficult cases.
Moreover, different quality requirements for image analysis algorithms may impact the prior workflow (e.g., optimization of coverslipping \cite{FerValCur2022, MatSch2022} or higher requirements for scanner calibration \cite{RomThoDex2022}), decreasing or increasing efforts in prior steps.
Algorithms are expected to make analyses more quantitative \cite{BerMcCByc2023}, which could lead to decreased inter-observer variability.
The assessment of complete concordance, however, could become more complex and time-consuming.

\subsubsection{Pitfalls}

Potential pitfalls on the technical level can be visible or silent.
Visible pitfalls include downtime of compute infrastructure or images rejected by image analysis algorithms.
Infrastructure downtime causes quantifiable delays and may require workarounds.
One can also measure an overall uptime or the frequency and duration of downtimes.
For images rejected by image analysis algorithms, the percentage of cases or slides affected and the percentage of affected diagnoses can be quantified.
In contrast, silent issues such as incorrect algorithmic results or ambiguous cases need to be detected by a human in subsequent reading or QC before such issues can be quantified in the same way.
Other pitfalls that may affect individual diagnoses and can be quantified on that basis are decision fatigue, i.e., the tendency to ignore frequently occurring warnings \cite{Kie2022} and automation bias, i.e., the over-reliance on algorithmic results \cite{Kie2022}, also caused by time pressure \cite{BerMcCByc2023}.

Perspectively, skills largely replaced by algorithms might be lost \cite{Kie2022}.
These effects can be mitigated by awareness and diligence.
Moreover, potential issues may arise, pertaining to usability of viewers.
Also, as mentioned above, algorithmic analyses may overload pathologists with information irrelevant for clinical care \cite{Kie2022}, or lack of interpretability of algorithmic results \cite{EvaRetGei2022}.
These potentials and pitfalls affect efficiency and work satisfaction in general but cannot be usefully evaluated on a per-case basis.
As for digital slide reading without advanced algorithm support, expectations of pathologists for algorithmic accuracy \cite{PatManGur2019} should be compared to observations.

All these effects, corresponding metrics, and potential mitigation approaches largely depend on which specific algorithmic solution(s) are employed in the workflow and how they are used (on demand, preprocessing with results visible to the reader, as `second read').
We are unaware of reported data which could be used for comparison.

\subsection{Overall Digitization Process and Change Management}\label{sec:ChangeManagement}

All stakeholder groups (Technicians, Pathologists, IT, Management, and Administration) need to be included for a digitization process to be successful \cite{CheAzhSng2016, SmiJohZeu2022, StaNguDie2020}.
A motivated project manager should drive the process \cite{EvgAleAnd2022}.
Pitfalls are insufficient technical expertise for the required decisions about components \cite{PatBalChe2021, JahPlaMoi2020} and potential conflicts of interest of external consultants (e.g., system vendors).
A trade-off is needed between involving enough persons so that no important aspects are missed and not involving too many persons to avoid excessive efforts and delays in decision-making.

\subsubsection{Technology}

A major general pitfall in technical decisions is selecting insufficiently modular and future-proof solutions, e.g., lack of support for the likely future standard of using the digital imaging and communications in medicine (DICOM) image format \cite{EscCarBoc2022}.
While possibly cheaper in the short term, this can lead to vendor lock-in \cite{HanArdReu2022, LeiBenMol2021, GooGilHar2013, ParPanSha2012} with reduced flexibility and increased costs in the future.
When switching to a new system, a major challenge is incorporating legacy data, which causes migration efforts for image data and metadata as well as ensuring data integrity \cite{DasJonMer2021}.
Also on the operational level, non-future-proof decisions can be problematic, e.g., when choosing image resolution or lossy (jpeg) compression only sufficient for current diagnostic purposes but not for future algorithmic image analysis \cite{StaVetHui2013}.
Another study reported that an initial choice of 1D barcodes allowed insufficient information density.
This resulted in an overlap of numbering ranges that cost 3 person-months to fix \cite{StaNguDie2020}.
Digitization in general, and computer-aided slide reading in particular, come with an increase in complexity and in the number of interacting hardware and software components.
This increases the potential for downtimes of single components affecting the entire workflow.
These issues can be mitigated by planning for sufficient redundancy to meet uptime goals based on estimated downtimes, possibly adjusted by observed data, and by aiming for robustness against isolated issues.

\subsubsection{Administrative Level}

On the administrative level, digitization of a lab involves procurement and installation efforts.
Details need to be negotiated, contracts signed, hardware and software delivered, and systems need to be installed, i.e., IT infrastructure, lab devices, and workstations.
Implementing IT infrastructure was reported to take a few days with prior experience \cite{FraCapGug2021}, and 60 IT person-hours at the hub site vs. 16 to 24 at spoke sites \cite{ChoPalFis2019}.
However, progress may be delayed due to procurement procedures, devices not being in stock, importing/customs procedures, personnel availability etc. \cite{MaySmeSem2022}, creating additional efforts and frustration when following up on work already agreed on.
This also applies to maintenance and repairs, where delays in delivery of replacement parts \cite{HarPanMcH2017} or no-visitor policies preventing on-site technical support \cite{ArdReuHam2021} causes delays.
Additional devices and changed workflows may also involve re-organizing rooms/work spaces \cite{FraCapGug2021}.
The efforts on the administrative level can also be seen as an investment if digitization simplifies new collaborations, e.g., due to fewer restrictions for digital data compared to physical slides \cite{ZhaWuxDin2015}.

\subsubsection{Operative Level}
On the operative level, effort is needed to get familiar with digital workflows (time spent on training and reduced productivity during familiarization) \cite{BelMetNau2013, LisBelBil2020}, which may take weeks to months \cite{RetAneMor2020b, Bai2015, RetAneMor2020a, FraCapGug2021, FerMonCoe2023}.
As workflows change for all lab personnel, training is needed for everyone involved \cite{MeaCarLor2017}; problems may arise if the adoption of digitization does not match the individuals' own pace \cite{Bai2015}.
Acceptance is likely higher if organizational changes are fewer rather than more \cite{Gau2022}, if they imply simplification \cite{Gau2022} and if the digitization process is communicated well \cite{FraLImAme2021, HarPanMcH2017}.
In particular, changes in lab workflows may also involve changes to quality assurance procedures \cite{RamTejUth2021}.
In any case, a diagnostic concordance analysis comparing previous and new workflows is needed \cite{RaoKumRaj2021, EvaBroBui2021, CroFurIga2018} and is helpful to become confident in digital slide reading \cite{WilTre2019}.
Changes in workflows also imply changes in working conditions, e.g., the possibility of working remotely.
However, only a few steps of the workflow can be performed remotely (2 of 11 \cite{ArdReuHam2021}), so this mostly benefits pathologists; not all staff will be able to work remotely.

\subsubsection{Personnel Efforts}

Moving to a digital workflow is typically a time-consuming iterative process with potential for iterative improvements: `It is completely normal not to get it with the first try' \cite{StaNguDie2020}.
Of the labs surveyed in \cite{PinBycTsu2023}, 51\% reported starting by digitizing selected cases, 41\% scan all histology routine cases or expect to do so within 2 years.
One lab reported a duration of about one year from planning to go-live \cite{Gau2022}, during which 1 to 1.5 full time equivalents (FTE) were involved: 0.4 lab project manager, 0.2 IT project manager, 0.1 to 0.6 LIMS expert, 0.2 physician, 0.1 technician \cite{Gau2022}.
Scanner evaluation was estimated to require 6 weeks for a technician and 40 hours each for a senior scientist and a pathologist \cite{ArdKleMan2023}.
There is typically a transition phase/cross-over period \cite{GriTre2017} with hybrid or parallel analog and digital workflows over a period of, e.g., 6 to 12 months \cite{StaNguDie2020} with temporarily increased costs \cite{ArdKleMan2023}.
Digitization can also be partial, using analog and digital workflows in parallel: `Benefits are evident also on lower percentages,' but a `complete digital workflow is the only viable option when a lab wants to transition to an AI-assisted workflow' \cite{StaNguDie2020}.

\subsection{Economic Performance Metrics}\label{sec:EconomicPerformanceMetrics}

The assessment of whether or not having digitized a lab is a success is based on whether the workflow has become better and/or cheaper, i.e., whether total turnaround times and/or costs have been reduced.
This addresses both the clinical \cite{WilBotTre2017} and the business \cite{LujQuiHar2021, WilBotCla2018} case for digitizing pathology labs.

\subsubsection{Turnaround Time}

The total turnaround time \cite{RaoKumRaj2021, HanReuSam2019, ChoPalFis2019} is a key performance indicator also used in derived metrics such as the percentage of cases with turnaround time below two days \cite{PatSmiKur2012, AlsAlS2015}.
Throughout the literature, the precise meaning of `turnaround' differs and may refer to different parts of the workflow.
By `total turnaround time,' we here refer to the duration of the entire workflow from accession to final sign-out.
Digitization affects the entire workflow; different requirements and equipment may increase or decrease convenience and processing time of individual steps \cite{FraCapGug2021}.
This impacts the total turnaround time, which is likely to decrease if time savings (transmission of digital data clearly faster than shipping/couriering physical slides, reduced handling efforts \cite{HoxAhlStr2014}) outweigh additional work (mainly scanning).

Total turnaround time is typically several days \cite{RaoKumRaj2021}.
Digitization can reduce this by up to one \cite{RaoKumRaj2021, StaNguSpo2019} or two \cite{VolChoEva2018, EvaVajHen2020} days.
Key changes in time needed in the workflow are that scanning takes additional time in digital workflows but saves the time needed for handling physical slides afterward, i.e., assembling cases and dispatching/couriering/mailing them.
In the case of teleconsultation, a reduction from 86 h to 35 min was reported \cite{HagTolSch2021} because no shipping was needed.
Removing the need for case assembly and delivery was reported to reduce the time from `staining completed' to `slide ready for pathologist' from 8h to 3h \cite{FraGarZan2017} and by 24 to 96 hours \cite{EvaVajHen2020}.
Electronic transmission of cases reduces the delay due to batch dispatching, e.g., saving on average one hour if sending cases by courier every two hours \cite{GuoBirFar2016}, as well as the effort of individually transporting urgent cases (e.g., 10\% of cases reported by \cite{BaiBucLee2018}).
Another study reported that their digital workflow took 3.8\% longer between accession and `slides ready,' but that the time from accession to final sign-out was 4.3\% faster \cite{ArdReuHam2021}.
For intraoperative consultations, replacing couriering slides by scanning and data transmission was reported to save 10 minutes \cite{ChaSadYao2022}, which is substantial in this use case.
For single cases, turnaround time can be reduced by prioritizing them over other cases \cite{ChoPalFis2019} (which, in turn, can be supported by digital workflows \cite{MayGooMen2023}).

Two further metrics related to turnaround time are the distance a specimen travels in the lab \cite{BlaDawHam2017} and the number of touch points in the workflow \cite{LujSavSha2021}.
The distance a specimen travels was investigated in the context of reorganizing a lab without digital slide reading \cite{BlaDawHam2017}, and decreases when physically transporting slides is replaced by digital data traveling automatically.
The number of touch points was minimized with the goal of minimizing staff interaction during the COVID-19 pandemic \cite{LujSavSha2021}, but can also reduce potential sources for mix-ups and delays.
Such delays can be aggravated by incompatible batch sizes between processing devices \cite{BlaDawHam2017}.

\subsubsection{Costs}

Changes in costs (additional investments and savings) due to digitizing a pathology lab can be categorized on a spectrum from one-time setup costs (e.g., installation and training), periodically occurring costs (e.g., scheduled maintenance, replacing hardware after end-of-life), and continuous costs for ongoing operations (e.g., shifts in personnel, consumables).
For many aspects, this distinction depends on the actual implementation (e.g., buying vs. leasing hardware; one-time software licenses vs. subscriptions/pay-per-case).
The costs consist of expenditures for procuring, e.g., hardware and software, work hours of external persons, and work hours of own personnel, all of which boil down to monetary amounts.
In addition, lost revenues due to potential delays should be considered.
Two challenges in the cost assessment are that additional costs are easier to calculate than financial gains \cite{BaiKhaLaa2023} and that not all aspects are easily convertible to monetary amounts, e.g., job satisfaction or attractiveness for employees in a competitive job market \cite{GroBlaHof2020, MatSch2022}, in particular for innovative faculty and staff \cite{ArdKleMan2023}, is not only a matter of salary \cite{LujQuiHar2021}.

Costs clearly depend on the size of the lab.
One study reported 1.3M USD saved in 5 years \cite{PatBalChe2021}, another studies reported 12.4M USD saved lab costs in 5 years plus 5.4M USD for avoided treatment (i.e., 16 USD per accession) \cite{HoxAhlStr2014}, in particular, 300k USD saved for not buying optical microscopes \cite{HoxAhlStr2014}; and 131k CAD saved in courier and travel-related costs \cite{EvaVajHen2020}.
Other studies observed an increase in process-cycle efficiency (as determined by value stream mapping) from 3\% to 57\% \cite{HagTolSch2021} or an increase of value per pathologist by 28\% \cite{RetAneMor2020a}.
Costs can be measured per time (i.e., month or year) or per patient/case/specimen/slide \cite{GenRemUnt2021}, which should be distinguished for meaningful interpretations and comparisons of more detailed assessments than an overall return-on-invest.

A breakdown of setup and operational costs \cite{EvgAleAnd2022, StaNguSpo2019, PatBalChe2021} in different ways helps better assess actual costs and compare them to prospective estimates.
A detailed cost analysis \cite{ArdKleMan2023} for a large laboratory found six major cost categories: scanner acquisition (annualized using 7-year depreciation) amounting to 21\% of the total cost; vendor scanner service (10\%); related IT infrastructure (annualized using 5-year depreciation; 33\%); initial staffing for the infrastructure stages (7\%); IT staff (10\%); and the digital scan team (21\%).
Costs per scan were estimated as at least 0.55 USD for scanner acquisition and maintenance and 0.79 to 0.92 USD for the digital scan team \cite{ArdKleMan2023}.

Changes in operational costs include changes in workforce, e.g., 6 fewer FTE pathologists, 2.5 fewer FTE histotechnologists, and three fewer FTE needed in the slide file room, but 3 FTE scanning technicians were needed for a total of 6 scanners \cite{RetAneMor2020a}.
A scan team of 21 individuals (at unspecified percentage) for 26 scanners was reported in \cite{ArdReuHam2021}.
For image QC, 1 FTE was estimated for every 3 to 4 WSI scanners \cite{ArdKleMan2023}.
Another study reported 5 minutes per case saved in lab logistics (3 min by digital handling and transmission, 1.5 min in preparation for multidisciplinary meetings), corresponding to 2.3 FTE lab staff \cite{BaiBucLee2018}.
For a hub--spoke setup of labs, substantial salary cost savings were reported for spoke sites, whereas the hub site does not expect to recompense system costs in the short term \cite{ChoPalFis2019}.
With a more complex technical setup and dependence on IT infrastructure in digitized labs \cite{JahPlaMoi2020}, the potential for unexpected costs should be considered.
Currently, specific reimbursement for digital diagnostic procedures \cite{HomLotSch2021} is very limited, so investments need to be compensated for by higher efficiency, i.e., reduced operational costs and/or higher throughput.

\section{Discussion}

The literature indicates that digitizing pathology labs has its challenges, but that it is possible to overcome the individual issues, and that the results are generally beneficial.
Pathologists are often skeptical at first, a completely normal human reaction when confronted with unfamiliar new tools, but happy with the results afterward.
Switching from analog microscopes to digital slide reading remains an investment with unclear immediate benefits, as reimbursement for digital diagnostic procedures is still limited \cite{HomLotSch2021}.
However, gains in efficiency through digitization will be necessary to meet the increasing diagnostic demand with a decreasing number of pathologists \cite{BabGunWal2021, BaiBucLee2018, BaiCakMer2016, BelMetNau2013, BroColRit2019, CroFurIga2018, GoaRanWil2017, GroBlaHof2020, Kie2022, KusRifRis2022, LaaLitCio2021, Lee2018a, LujQuiHar2021}.

\begin{table}[t!]
  \centering
  \caption{Overview of metrics for assessing the implications of digitizing a pathology lab and which purposes they are suited for.
    For comparison over time with varying case load and for comparison between labs, numbers should be appropriately normalized.}\label{tab:metricsOverview}

  \renewcommand*{\baselinestretch}{1.2}\small
  \begin{tabular}{>{\raggedright}p{0.07\textwidth} >{\raggedright}p{0.34\textwidth} >{\raggedright}p{0.34\textwidth} >{\raggedright}p{0.14\textwidth}}
    \toprule
                                       & Metrics                                                                                                             & Purpose                                                 & See Results Subsections(s)                                                                                                          \tabularnewline
    \midrule
    Devices                            & Device specification (e.g., storage capacity, display size and resolution)                                          & Comparison to requirements                              & \multirow[t]{3}{=}{\nameref{sec:WholeSlideScanning}, \hbox{\nameref{sec:ITInfrastructure}}, \nameref{sec:PathologistsWorkstations}} \tabularnewline
                                       & Processing capacity: specimens/slides per hour/day (considering peak load)                                          & Choosing models and number of devices, comparison observed in routine vs. specified/required                                                                                                  \tabularnewline
                                       & Downtimes per day/month; as percentage of time or minutes per day/month (considering multiplicate equipment)        & Internal communication and planning; planning redundancy, comparison observed in routine vs. specified/required                                                                               \tabularnewline
    \midrule
    Processing                         & Individual processing issues: percentage of slides affected                                                         & Monitoring, optimization of individual processing steps & \multirow[t]{4}{=}{\nameref{sec:SpecimenHandling}, \nameref{sec:WholeSlideScanning}, \nameref{sec:DigitalSlideReading}}             \tabularnewline
                                       & Diagnoses affected by specific issues: percentage of cases                                                          & Prioritization of issues, quality monitoring                                                                                                                                                  \tabularnewline
                                       & Slides/cases affected by issues per day/month                                                                       & Internal communication and planning                                                                                                                                                           \tabularnewline
                                       & Processing delays (time) due to issues/device downtimes                                                             & Quality monitoring                                                                                                                                                                            \tabularnewline
    \midrule
    \multirow[t]{3}{=}{Result Quality} & Concordance of diagnoses: percentage                                                                                & Establish non-inferiority of digital workflow           & \multirow[t]{2}{=}{\nameref{sec:ConcordanceAnalysis}}                                                                               \tabularnewline
                                       & Quantification/Classification accuracy (depends on task, e.g., correlation coefficient, Cohen's κ, AUROC, F1 score) & Assessment of algorithms                                                                                                                                                                      \tabularnewline
    \midrule
    Economic Performance               & Efforts (work, consumables) for lab operation and for resolving issues: monetary amount per month; per case         & Financial planning; cost monitoring                     & \multirow[t]{3}{=}{\nameref{sec:EconomicPerformanceMetrics}}                                                                        \tabularnewline
                                       & Work and automatic processing times per case                                                                        & Total turnaround time                                                                                                                                                                         \tabularnewline
                                       & Job satisfaction (via survey; qualitative or semi-quantitative)                                                     & Maximizing staff retention                                                                                                                                                                    \tabularnewline
    \bottomrule
  \end{tabular}
\end{table}

Metrics assessing the implications of digitizing pathology labs are indispensable for planning and monitoring successful implementation and operation.
For a meaningful interpretation of the data and comparison to other labs, an appropriate normalization for the specific assessment purpose is essential, see Table \ref{tab:metricsOverview}.
Some aspects are hard to quantify in a meaningful way, such as ergonomics, device/software usability, and job satisfaction.
They can be monitored by surveys, but numerical comparisons may be misleading due to subjective interpretation of response scales.

\subsection{Limitations}

The numbers reported here need to be interpreted with substantial care.
Meaningful quantitative comparisons between labs can be hard.
First, not all effects and aspects are quantifiable in a meaningful way (e.g., convenience, job satisfaction).
Environments of labs differ, e.g., whether the lab is part of a hospital or receives cases from different hospitals, typical cases, and funding (reimbursement differs between countries \cite{HomLotSch2021}).
In particular, the published experience reports compare different status quo ante to different outcomes of the digitization process.
Reporting is heterogeneous in terms of aspects addressed, metrics (per case/specimen/slide/diagnosis or per shift/day/month/year, which may be nontrivial to convert), and descriptive statistics used (e.g., mean vs. median).
Also, the terminology to describe issues is heterogeneous throughout the literature, requiring additional care when drawing comparisons.

\subsubsection{Biases}

Data reported in the literature is likely subject to biases.
On the one hand, issues and corresponding percentages are only reported by those monitoring the issues, e.g., only 29\% of labs surveyed in \cite{PinBycTsu2023} record scan error rates.
This may lead to a `complaint bias', exaggerating the extent of issues when combining findings from multiple studies.
On the other hand, success stories and, specifically, results of concordance analyses may be subject to publication bias overemphasizing successes.
However, the substantial number of publications reporting observed issues does not indicate the presence of major publication bias.
Moreover, software or device manufacturers being involved in publications might lead to bias in the reported information.
These biases are also reflected in the present overview, albeit without data on the extent and impact of such biases.
This overview is furthermore biased by the availability of information from labs willing to share their experience and spend the effort of writing publications.
A number of conference presentations (e.g., \cite{Ard2022, FukByc2022, Kli2022, MonZarAre2022, SanSoc2022, SidMatJos2022}) and other formats (e.g., webinars such as \cite{Gau2022, Byc2020}) present further experience.
However, this information is not readily available for creating an overview as presented here.
Costs are subject to the bias that many reporting laboratories have not fully transitioned to digital workflows so that savings due to discontinued steps no longer necessary have not yet materialized.

\subsubsection{Missing Data}

Some important aspects are scarcely represented in the existing literature.
Many studies focus on working conditions for pathologists \cite{MaySmeSem2022, PatBalChe2021, HanReuSam2019, HanReuArd2020, RetAneMor2020b, HanReuHam2019, DroMilVos2022, KelSolGar2020}; the perspective of other groups of lab staff \cite{SmiJohZeu2022, GarKunKel2020} is poorly represented.
As computer support for diagnosis only recently became available for routine use, real-world experience is scarce \cite{SanSoc2022}; time will tell which additional pitfalls may arise.Image analysis for computer-aided slide reading will make diagnoses more quantitative \cite{BerMcCByc2023}.
This offers the possibility to compute additional metrics, necessitating the choice of meaningful metrics for drawing useful and actionable conclusions.
Moreover, suitable metrics will be needed to verify the accuracy of quantitative results of algorithmic solutions \cite{RomThoDex2022}.

\subsubsection{Broader Perspective}

Working with medical data, IT security is a relevant topic when digitizing pathology labs.
Independent labs need to take care of this on their own, whereas labs at hospitals need to address this issue as part of a larger organization.
Our literature search revealed no data on security breaches (number of incidents per time, number of affected patients, cost of recovering from cyberattacks), possibly due to a reporting bias.
Furthermore, we (and the cited literature) here consider costs at the lab level, neglecting the bigger picture: cost for hospitals (including other diagnostic approaches, treatment costs, time saved in tumor boards, etc.) can be addressed by decisions at the hospital level; costs for the healthcare system at large can be addressed at the political level.

\subsubsection{Prospective Assessment}

Our retrospective view focuses on assessing the process after it has been decided and implemented.
Partially, this data can also be used for prospective estimates and for making decisions.
In this case, the available comparison between `before' and `after' digitization needs to be translated into estimates for the respective lab planning to digitize.
This means taking into account changing diagnostic demand and available workforce (in particular pathologists) and comparing estimates of `with digitization in three years' to `without digitization in three years' rather than to `now.' Moreover, decisions based on metrics should consider whether the respective metrics have become targets and thus ceased to be useful metrics.

\subsection{Conclusion}

Digitization of pathology labs disrupts established processes in many ways and is therefore inherently challenging.
Taking into account the lessons learned by other labs helps one to implement the changes in less time and with less effort and to avoid common pitfalls.
Similarly, the metrics described in this paper help in planning and monitoring successful implementation and operation.

\section{Author Contributions (CRediT)}

Conceptualization: AH, LOS.
Investigation: LOS
Writing---original draft: LOS, TRK.
Writing---review and editing: All authors.
Funding acquisition: AH, NZ.

\section{Funding}

This work was supported by the German Federal Ministry for Economic Affairs and Climate Action (BMWK) via the EMPAIA project (grant numbers 01MK20002A [NZ, RC, TRK] and 01MK20002B [AH, LOS]).

\section{Conflicts of Interest}

None.

\section{Data Availability Statement}

Data sharing is not applicable to this article as no datasets were generated or analyzed during the current study.

\interlinepenalty10000
\bibliography{literature}

\end{document}